\documentclass[twocolumn,prb,showpacs]{revtex4}

\usepackage{graphicx}
\usepackage{rotating}
\usepackage{amsmath}
\usepackage{amsfonts}
\usepackage{amssymb}
\usepackage{enumerate}
\usepackage{longtable}
\usepackage{dcolumn}
\usepackage{bbm}

\begin{document}

\title{Optimized Dynamical Decoupling for Time Dependent Hamiltonians}

\author{Stefano Pasini}
\email{pasini@fkt.physik.tu-dortmund.de}
\affiliation{Lehrstuhl f\"{u}r Theoretische Physik I,
Technische  Universit\"{a}t Dortmund,
 Otto-Hahn Stra\ss{}e 4, 44221 Dortmund, Germany}

\author{G\"otz S. Uhrig}
\email{goetz.uhrig@tu-dortmund.de}
\affiliation{Lehrstuhl f\"{u}r Theoretische Physik I,
Technische  Universit\"{a}t Dortmund,
 Otto-Hahn Stra\ss{}e 4, 44221 Dortmund, Germany}

\date{\rm\today}

\begin{abstract}
The validity of optimized dynamical decoupling (DD) is extended to
analytically time dependent Hamiltonians. 
As long as an expansion in time is possible the time dependence of the 
initial Hamiltonian does not affect the efficiency of optimized dynamical 
decoupling (UDD, Uhrig DD). This extension provides the
analytic basis  for (i) applying UDD  to effective Hamiltonians
in time dependent reference frames, for instance in the interaction
picture of fast modes and  for (ii) its application in hierarchical
 DD schemes with $\pi$ pulses about two perpendicular axes in spin space.
to suppress general decoherence, i.e., longitudinal relaxation
and dephasing.
\end{abstract}



\maketitle

\section{Introduction}

Progress in quantum information processing (QIP) requires a complete and 
coherent control of the dynamics of a quantum bit (spin $S=1/2$) coupled to an 
environment (bath). In particular, one must be able
to realize the no-operation reliably and coherently 
for long-time storage of quantum memory.
Hence decoherence must be suppressed. The most general 
decoherence  consists of both transversal dephasing and 
longitudinal relaxation, i.e., the decoherence rates
$1/T_2^\star$ and $1/T_1$, respectively, in nuclear magnetic resonance
(NMR) language. 

So far, only models without explicit time dependence have been considered
to our knowledge. The terms in the Hamiltonian $\widetilde H$
without coherent control
(we will call this Hamiltonian henceforth the initial one)
do not have any explicit dependence on the time. For such a model 
techniques of various degrees of sophistication 
exist to suppress the dephasing and/or the relaxation \cite{haebe76,freem98}. 
We concentrate here on the dynamical decoupling (DD) 
\cite{viola98,ban98,viola99a,khodj05,uhrig07,lee08a,uhrig08,yang08,uhrig09b} 
which generalizes the original ideas on spin echo techniques to 
open systems and their application to QIP \cite{hahn50,carr54,meibo58}.
Intuitively, the interaction  between spin (qubit) and bath is averaged to 
zero by means of repetitive $\pi$ pulses. Each pulse rotates the spin by an 
angle $\pi$ about a spin axis $\hat a$, thus inverting its 
components perpendicular to $\hat a$.

A particularly efficient way to suppress pure dephasing
is the optimized DD (Uhrig DD) 
\cite{uhrig07,lee08a,uhrig08,yang08,bierc09a,bierc09b,du09}
 where the instants $t_j$ ($j\in\{1,2\ldots N\}$) at which
$N$ instantaneous $\pi$ pulses \footnote{In the present work we stick
to  idealized, instantaneous pulses, but this
requirement can be released \cite{uhrig09c}.}
are applied are given by $t_j=T\delta_j$ where $T$ is the
total time of the sequence and
\begin{equation}
\label{udd}
\delta_j =\sin^2(j\pi/(2N+2)).
\end{equation}
By efficient suppression it is meant that each pulse
helps to suppress dephasing in one additional order in an
expansion in $T$, i.e., $N$ pulses reduce dephasing
to ${\cal O}(T^{N+1})$. The existence of an 
expansion in powers of $T$, at least as an asymptotic expansion, is
a necessary assumption.

So far, the derivation of the properties of UDD as defined in
Eq.\ \eqref{udd}
was given for time independent initial Hamiltonians 
\cite{uhrig07,lee08a,uhrig08,yang08}. The present study
extends this derivation to initial Hamiltonians including
an analytic time dependence. This extension is a breakthrough
because it establishes the applicability of 
optimized DD for effective Hamiltonians in special reference
frames, e.g., rotating frames, which induce an \emph{explicit} 
time dependence. Such situations arise also where fast modes are treated
in the interaction picture, are averaged over, or  integrated out
so that time dependent actions result
and these actions are sufficiently smooth in time.
The condition on smoothness need not always be fulfilled.

Another important application of UDD for time dependent Hamiltonians
is the suppression of general decoherence by the application
of $\pi$ pulses around two perpendicular spin axes on two
hierarchical levels. If UDD worked only for time independent 
initial Hamiltonians the only known solution for the secondary level 
would be concatenation of primary UDD sequences \cite{uhrig09b}. 
But recent numerical data by West {\it et al.} showed that 
also the suppression on the secondary level can be efficiently
realized by UDD \cite{west09}. They 
called the scheme quadratic DD (QDD). Hence the derivation below provides
the analytic foundation for the applicability of QDD.

For the sake of simplicity, we first give the extended derivation
for pure dephasing, addressing longitudinal relaxation
in a second step. Then the applications are discussed again
and we provide an explicit derivation of the time dependence
of the effective Hamiltonian after the primary application of
UDD.

\section{Dephasing}
We consider the explicitly time dependent Hamiltonian
\begin{equation}
\label{hamil1}
\widetilde H(t) =H_\mathrm{b}(t) + \sigma_z A_z(t)
\end{equation}
is studied
where the time dependences of the bath Hamiltonian $H_\mathrm{b}(t)$ and
of the coupling operator $A_z(t)$ are required to be analytic, i.e.,
they can be expanded in $t$. If the system described by \eqref{hamil1}
is subject to $N$ instantaneous $\pi$ pulses at the
instants $\{T\delta_j \}, j\in\{1,2,\ldots,N\}$ about a spin axis 
perpendicular to the $z$-axis the effective Hamiltonian $H(t)$ in the basis
of unflipped spins reads
\begin{equation}
\label{hamil2}
H(t) =H_\mathrm{b}(t) + \sigma_z A_z(t)F(t)
\end{equation}
where the switching function $F(t)=\pm1$
appears which changes sign at the instants $\{T\delta_j\}$. We are 
interested in the time evolution operator $U(t)$ induced by $H(t)$ 
\begin{subequations}
\begin{eqnarray}
\label{Udef}
U(T) &:=& {\cal T}\exp\big(
-i\int_0^T H(t)dt
\big)
\\
\label{Uinter}
 &=& U_0(T)U_1(T) 
\end{eqnarray}
\end{subequations}
where ${\cal T}$ is the time-ordering operator.
The second line is based on the interaction picture
with respect to $H_\mathrm{b}(t)$
\begin{subequations}
\begin{eqnarray}
\label{U0def}
U_0(t) &:=& {\cal T}\exp\big(
-i\int_0^T H_\mathrm{b}(t)dt
\big)
\\
\label{U1def}
U_1(t) &:=& {\cal T}\exp\big(
-i\int_0^T A_\mathrm{I}(t)F(t) dt
\big)
\\
A_\mathrm{I}(t) &:=& U_0^\dagger(t)\sigma_z A_z(t)U_0(t)\ .
\end{eqnarray}
\end{subequations}
The key observation is that $A_\mathrm{I}(t)$ is analytic
as well because $H_\mathrm{b}(t)$ is analytic according
to our requirement, and thus $U_0(t)$, and $A_z(t)$ 
again according to our requirement \footnote{We assume 
that all operators appearing in the derivation
are bounded. Alternatively, we require at least that 
all products of operators of the system have
finite matrix elements. This implies that no divergences at low or
high energies are present \cite{cywin08,uhrig08,pasin10a}.}.
 Thus we have
\begin{equation}
\label{Aseries}
A_\mathrm{I}(t) =\sum_{p=0}^\infty A_p t^p.
\end{equation}
To be precise, to exclude any terms up to a given order $N$
we only need that $A_\mathrm{I}(t)$ can be represented
by  the sum in \eqref{Aseries} up to $p=N$ plus a
residual function of higher order, i.e., the convergence
of the Taylor series is not needed.
In general, the operators $A_p$ are complicated
integral expressions of the operators in \eqref{hamil1}.

Next, the time evolution $U_1(T)$ is expanded 
according to standard time dependent perturbation theory
\begin{subequations}
\begin{eqnarray}
U_1(t) &=& \sum_{n=0}^\infty (-i)^n u_n
\\
u_n &=& \int_0^T F(t_n)
\ldots  \int_0^{t_3}  F(t_{2}) \int_0^{t_2}  F(t_{1})
\nonumber
\\
&\times&A_\mathrm{I}(t_n)  A_\mathrm{I}(t_{n-1}) \ldots 
 A_\mathrm{I}(t_{1}) dt_1dt_2 \ldots  dt_n.
\qquad
\label{un_def}
\end{eqnarray}
\end{subequations}
Our aim is to show that the powers with $n$ odd
are of order $T^{N+1}$ because only the odd powers
in $\sigma_z$ affect the qubit spin. Hence
we can follow the reasoning of Yang and Liu  \cite{yang08}
from here on. In order to keep the present communication
self-contained we include the main steps. First, we
 expand in powers of $T$ by inserting
\eqref{Aseries} into \eqref{un_def}
\begin{equation}
u_n=\sum_{\{p_j\}} T^{n+P_n}
A_{p_n}\ldots A_{p_2} A_{p_1} F_{p_1,p_2,\ldots p_n}
\end{equation}
where $p_j\in\mathbb{N}$ and $P_n:=\sum_{j=1}^n p_j$
and
\begin{equation}
\label{Fcoeff}
F_{p_1\ldots p_n} :=
\int_0^{1} d\tilde t_n  \ldots \int_0^{\tilde t_3} d\tilde t_{2}
\int_0^{\tilde t_2} d\tilde t_1 \prod_{j=1}^n F(T\tilde t) \tilde t_j^{p_j}.
\end{equation}
We used the dimensionless relative times $\tilde t:=t/T$. Since
the $N$ switching instants are given by $T \delta_j$
the function $F(T\tilde t)$ does not depend on $T$ for given $\{\delta_j \}$.
Hence the coefficients $F_{p_1\ldots p_n}$ do not depend on $T$.

Our goal is to show that $F_{p_1\ldots p_n}$
vanishes for $n$ odd and $N\ge n+P_n$. Based on the UDD choice
for the $\{ \delta_j \}$ in \eqref{udd} the substitution 
$\tilde t =\sin^2(\theta/2)$ suggests itself because it 
renders $f(\theta):=F(T\sin^2(\theta/2))$ particularly 
simple if the $\{ \delta_j \}$ are chosen according to \eqref{udd}.
Then $f(\theta)=(-1)^j$ holds for $\theta\in
(j\pi/(N+1),(j+1)\pi/(N+1)$ with $j\in\{0,\ldots,N\}$.
If we release this contraint on $j$ allowing $j\in\mathbb{Z}$
the function $f(\theta)$
becomes an odd function with antiperiod $\pi/(N+1)$.
Thus its Fourier series 
\begin{equation}
\label{fourier}
f(\theta) = \sum_{k=0}^\infty c_{2k+1} \sin((2k+1)(N+1)\theta)
\end{equation}
contains only harmonics
$\sin(r\theta)$ with $r$ an odd multiple of
$N+1$. The precise coefficients $c_{2k+1}$
do not matter which can be exploited 
for other purposes, e.g., to deal with
pulses of finite duration \cite{uhrig09c}.

Under the substitution $\tilde t =\sin^2(\theta/2)$
the terms $\tilde t^p dt$ in \eqref{Fcoeff}
become $\sin^{2p}(\theta/2)\sin(\theta)d\theta$
which can be reexpressed as suitably weighted sum over
terms $\sin(q\theta)d\theta$ with $q\in\mathbb{Z}, |q|\le p+1$.
Thus we achieve our goal if we can show that the coefficients
\begin{equation}
\label{fcoeff}
f_{q_1\ldots q_n} :=
\int_0^{\pi} d \theta_n  \ldots \int_0^{\theta_3} d\theta_{2}
\int_0^{\theta_2} d\theta_1 \prod_{j=1}^n f(\theta_j)
\sin(q_j\theta_j)
\end{equation}
vanish for $n$ odd and $|q_j|\le p_j+1$. These coefficients
are split up further by inserting the Fourier series \eqref{fourier}
for $f(\theta)$ consisting of terms $\sin(r\theta)$ ($r$
odd multiple of $N+1$).
Twice the product of two sine functions
is the difference of two cosines
whose arguments are sum and difference of the sine arguments.
Thus we want to show
\begin{equation}
\label{recurs-start}
0=\int_0^{\pi} d \theta_n  \ldots \int_0^{\theta_3} d\theta_{2}
\int_0^{\theta_2} d\theta_1 \prod_{j=1}^n \cos((r_j+q_j)\theta_j)
\end{equation}
where $r_j$ is an odd multiple  of
$N+1$ and 
\begin{equation}
\sum_{j=0}^n |q_j| \le \sum_{j=0}^n (p_j+1) =n+P_n \le N.
\end{equation}
It is easy to perform
the first two integrations in \eqref{recurs-start}
\begin{equation}
\cos((r_3+q_3)\theta_j)
\int_0^{\theta_3} d\theta_{2}
\int_0^{\theta_2} d\theta_1 \prod_{j=1}^2 \cos((r_j+q_j)\theta_j)
\end{equation}
analytically yielding a lengthy sum over terms of the form
$\cos((r'_3+q'_3)\theta_3)$
where $r'_3$ is still an odd multiple  of
$N+1$ and $|q'_3|\le |q_1|+|q_2|+ |q_3|$.
Hence the structure of the expression on the
r.h.s.\ of \eqref{recurs-start} is preserved, but $n$ is lowered
by two. This procedure is iterated till $n=1$ and we
arrive at
\begin{equation}
\label{recurs-end}
0=\int_0^{\pi} d \theta \cos((R+Q)\theta)
\end{equation}
because $R$ is an odd multiple  of
$N+1$ and $|Q|\le N$, thus $|R+Q|\in\mathbb{N}$. This concludes the 
derivation.

\section{Longitudinal relaxation}
The above derivation
holds also for the odd powers of longitudinal relaxation
as observed for constant Hamiltonians before \cite{yang08}.
The Hamiltonian studied is
$\widetilde H(t) = D_0(t) + D_1(t)$ where
\begin{subequations}
\label{DD}
\begin{eqnarray}
\label{D0}
D_0(t)  &:=& H_\mathrm{b}(t) + \sigma_z A_z(t)
\\
\label{D1}
D_1(t) &=& \vec\sigma_\perp\cdot \vec A_\perp(t)=
\sigma_x A_x(t) +\sigma_y A_y(t).
\end{eqnarray}
\end{subequations}
The $\pi$ pulses are applied around the spin $z$ axis so that
the switching function appears for $D_1$, i.e., 
$H(t) = D_0(t) + D_1(t)F(t)$. Then all the above steps for dephasing
can be repeated identically on substituting $H_\mathrm{b} \to D_0$
and $\sigma_z A_z\to D_1$. Thus we know that the UDD sequence 
of $N$ pulses  makes all odd powers up to $N$ in $D_1$ vanish.
Thus it efficiently suppresses longitudinal relaxation also
for time dependent Hamiltonians.

\section{Applications}
The first application, of course, is the use of the UDD
for Hamiltonians in a certain reference frames which imply
that the Hamiltonians are effective ones with some time
dependence. Examples are a rotating frame, in which a magnetic
field is not fully compensated, or an effective Hamiltonian
in which fast modes have been averaged by the help
of Magnus expansions or they are treated in an interaction picture. 
In the cases, where
 the resulting time dependence can be considered to be sufficiently smooth
(which need not be always true)
the previous results establish the applicability of the
optimized sequence UDD, in spite of the time dependence of
the Hamiltonian.

\begin{figure}[ht]
    \begin{center}
    \includegraphics[width=0.99\columnwidth,clip]{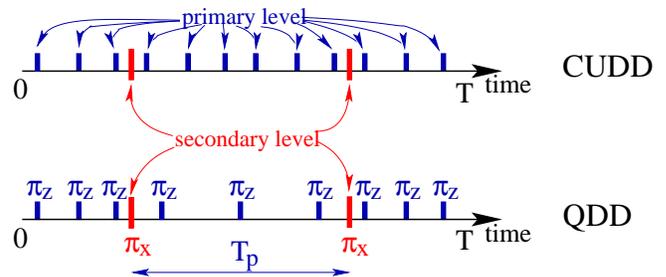}
    \end{center}
    \caption{(Color online) 
      Examples of nested DD suppressing general decoherence. 
      Here the primary $\pi$ pulses rotate about the $z$ axis and the 
      secondary ones about $x$.
      But any pair of perpendicular axes will do \cite{endnote25}. 
      In practice,
      the primary level should compensate the faster decoherence.
      The duration of the primary sequences is $T_\mathrm{p}$, the
      total duration $T$.
      \label{fig:nesting}
}
\end{figure}

The second application concerns dynamic decoupling for general 
decoherence. The nesting of pulse sequences 
about  perpendicular spin axes makes it possible
to eliminate all possible couplings between a qubit
and its environment  provided the expansion in time
is possible, see Fig.\ \ref{fig:nesting}. A first
proposal used iterative concatenation (CDD) without optimization
leading to a fast growing number of pulses proportional to $4^\ell$
if terms up to $T^\ell$ should be eliminated \cite{khodj05}.
If one uses concatenation on the secondary level, but optimized
UDD on the primary level (i.e.\ CUDD)
the number of pulses grows only like $2^\ell$ improving
by a square root \cite{uhrig09b}.
On the primary level, CUDD suppresses longitudinal
relaxation by $N_z$ $\pi$ pulses about the $z$ axis
such that the time evolution due to a Hamiltonian 
$\widetilde H = D_0 + D_1$ (for $D_i$ see
(\ref{DD})) without explicit time dependence is
reduced to the time evolution due to an
effective Hamiltonian
\begin{equation}
\label{heff}
\widetilde H^\mathrm{eff}(t) = 
D_0^\mathrm{eff}(t)
\end{equation}
up to correction of the order 
${\cal O}(T^{N_z+1}_\mathrm{p})$. 
For a discussion of the smoothness of  $\widetilde H^\mathrm{eff}(t)$
we refer the reader to the Appendix.
The time dependence of $\widetilde H^\mathrm{eff}(t)$
 was seen previously as the decisive
obstacle to apply  an optimized UDD sequence again on the secondary level
\cite{uhrig09b}. 
Below we show explicitly that $H^\mathrm{eff}(t)$
is indeed time dependent.

Very recent numerical data \cite{west09}, however,
indicates that the use of a UDD sequence also on
the secondary level with pulses about $z_\perp$ is in fact a very efficient
way to suppress general decoherence. This nesting of
two perpendicular UDD sequences of $N_z$ $\pi$ pulses about
$z$ for each of the $N_\perp+1$ intervals of a UDD sequence
of $N_\perp$ $\pi$ pulses about a perpendicular axis is
called quadratic dynamic decoupling (QDD) \cite{west09}
highlighting the case $N_z=N_\perp$. This choice is advantageous if
longitudinal relaxation and dephasing are of similar
magnitude.

The above derivation of the UDD as optimized sequence
for dephasing and longitudinal relaxation for \emph{time dependent}
Hamiltonians provides the analytic foundation of the
QDD proposed by West {\it et al.} \cite{west09}.
In spite of the time dependence that the effective
Hamiltonian $\widetilde H^\mathrm{eff}(t)$
acquires under the primary UDD (suppressing
longitudinal relaxation up to $T^{N_z+1}$) the secondary UDD (suppressing
dephasing up to $T^{N_\perp+1}$)
will still work to the desired order given by
the number of pulses \footnote{One can equally suppress
dephasing on the primary level and relaxation on the
secondary level.}. The reason is that an analytic
time dependence, quite surprisingly, does not spoil
the analytic properties of the optimized UDD sequences.

\section{Time dependence after the primary DD}
Applying $\pi$ pulses about the spin axis $z$ to 
 $\widetilde H = D_0 + D_1$ \emph{without}
explicit time dependence converts $\widetilde H$
to $H(t) = D_0 + D_1 F(t)$ where $F(t)=\pm1$ is
switching sign at the instants
of the pulses \cite{endnote23}, see Fig.\ \ref{fig:nesting}. To
 find the time dependence of the time evolution $U(T_\mathrm{p})$
due to $H(t)$ the Magnus expansion 
$U(T_\mathrm{p})=\exp\{-i T_\mathrm{p} 
\left(H^{(1)}+H^{(2)}+H^{(3)}\right)+{\cal O}(T_\mathrm{p}^4)\}$
is used \cite{magnu54,haebe76}.
 The terms are powers in $T_\mathrm{p}$; they read
$T_\mathrm{p} H^{(1)}=\int_0^{T_\mathrm{p}} {H}(t)\mathrm{d}t$, 
$T_\mathrm{p} H^{(2)}=i/2\int_0^{T_\mathrm{p}}\int_0^{t_1} 
[{H}(t_1),{H}(t_2)]\mathrm{d}t_1\mathrm{d}t_2$  and
\begin{eqnarray}
 \label{eq:magnus_H3}
 T_\mathrm{p} H^{(3)}=&-&\frac{1}{6}\int_0^{T_\mathrm{p}}
\int_0^{t_1}\int_0^{t_2} 
 \left\{[{H}(t_1),[{H}(t_2),{H}(t_3)]]\right.
\nonumber \\
&+&\left.[{H}(t_3),[{H}(t_2),{H}(t_1)]]\right\}
\mathrm{d}t_1\mathrm{d}t_2\mathrm{d}t_3.
\end{eqnarray}

The first order term $H^{(1)}$
contains the integral $I_1:=\int_0^{T_\mathrm{p}} F(t)\mathrm d t$
which  vanishes for any reasonable DD sequence with
at least one pulse, thus one has $H^{(1)}=D_0$
without time dependence.

The second order term $H^{(2)}$
involves the commutator $[{H}(t_1),{H}(t_2)]$. We find
\begin{equation}
 \label{eq:magnus2} 
T_\mathrm{p} H^{(2)}=\frac{1}{2}\hat{\eta}^{(2)}\int_0^{T_\mathrm{p}}
\int_0^{t_1} 
\left(F(t_2)-F(t_1)\right)\mathrm{d}t_1\mathrm{d}t_2, 
\end{equation} 
where the operator $\hat{\eta}^{(2)}$ induces dephasing 
\begin{eqnarray}
 \label{eq:eta2}
\hat{\eta}^{(2)}&:=& 
\sum_{i,j=x,y}\sigma_i \left([H_\mathrm{b},A_i]
-i\epsilon^{zij}\{A_z,A_j\}\right),
\end{eqnarray} 
with $\epsilon^{ijz}$ being the Levi-Civita operator and $\{,\}$ the
 anti-commutator. Since the r.h.s.\ of \eqref{eq:magnus2}
is linear in $F(t)$ we know from the general derivations 
given above or in Ref.\ \onlinecite{yang08} that the double
integral vanishes if a UDD sequence with 
two or more pulses is applied since the total order of the
term is $T_\mathrm{p}^2$ \footnote{For symmetric pulse sequences, e.g.,
UDD sequences with an even number of pulses, a general theorem
on the Magnus expansion precludes a finite second order term
anyway \cite{haebe76}; note our non-standard
counting of the orders.}.

The third order finally yields a non-vanishing
time dependence. There is a  linear  and a quadratic 
term in $F(t)$
\begin{subequations}
\begin{eqnarray}
 \label{eq:magnus3}
T_\mathrm{p} H^{(3)}&=&-\frac{1}{6}\left\{I_{3,1}[D_0,\hat{\eta}^{(2)}]
+I_{3,2}[D_1,\hat{\eta}^{(2)}]\right\}
\qquad
\end{eqnarray}
\begin{equation}
 \label{eq:I31} 
I_{3,1}:=\int_0^{T_\mathrm{p}}\int_0^{t_1}\int_0^{t_2}
\left(F(t_3)+F(t_1)-2F(t_2)\right) 
\mathrm{d}t_1\mathrm{d}t_2\mathrm{d}t_3
\end{equation}
\begin{eqnarray}
 \label{eq:I32} 
I_{3,2}&:=&\int_0^{T_\mathrm{p}}\int_0^{t_1}\int_0^{t_2}
\left(2F(t_1)F(t_3)\right.
\nonumber \\
&-&\left.F(t_2)F(t_1)+F(t_2)F(t_3))\right) 
\mathrm{d}t_1\mathrm{d}t_2\mathrm{d}t_3.\qquad
\end{eqnarray}
\end{subequations} 
We do not give the first commutator in (\ref{eq:magnus3}) 
explicitly because $I_{3,1}$ is linear in $F(t)$. Hence it is zero
for any UDD with three or more pulses. But  $I_{3,2}$
is quadratic in $F(t)$ so that the UDD sequence
does not make any statement on its value. Hence it will
generally be finite. Indeed, we verify numerically for $N_z$ pulses
the examples  $I_{3,2}^{N_z=3}=-0.03033T_\mathrm{p}^3$ and 
$I_{3,2}^{N_z=7}=-0.00668T_\mathrm{p}^3$ for $N_z$ odd and 
$I_{3,2}^{N_z=6}=-0.00884T_\mathrm{p}^3$ and 
$I_{3,2}^{N_z=8}=-0.00524T_\mathrm{p}^3$ for $N_z$ even.
This implies a finite quadratic dependence of the effective
Hamiltonian $H^\mathrm{eff}$ which we aimed to establish.

For completeness we compute the corresponding operator 
$[D_1,\hat{\eta}^{(2)}]$ which introduces corrections to dephasing
and to the bath dynamics as expected
\begin{eqnarray}
 \label{eq:comm2_magnus3} 
[D_1,\hat{\eta}^{(2)}]&=& 
 \sum_{i,j=x,y}\left(\left[A_i,[H_\mathrm{b},A_i]
-i\epsilon^{ijz}\{A_z,A_j\}\right]\right.
\nonumber\\
&+&i\sigma_z\left.\left\{A_i,\epsilon^{ijz}[H_\mathrm{b},A_j]+
i\{A_z,A_i\}\right\}\right).\quad
\end{eqnarray}  
Obviously, these terms do not vanish except for very special choices
of bath $H_\mathrm{b}$ and coupling operator $\vec A$. This
 completes the derivation that the effective Hamiltonians
are indeed time dependent. Thus the analytic argument in Ref.\ 
\onlinecite{west09} for the validity of the QDD does not
hold.

\section{Summary}
We extended the derivation of the optimized properties
of Uhrig dynamic decoupling to initial Hamiltonians with time dependence.
First, this establishes the applicability of optimized dynamic decoupling
also for the large class of effective Hamiltonians which inherit
an explicit time dependence from a special reference
frames or from the treatment of fast modes
in the interaction picture, by average Hamiltonian theory, or 
by integrating them out.
Second, our finding provides the analytic reason for the 
advantageous properties of  quadratic
dynamic decoupling for the suppression of general decoherence
including both dephasing and longitudinal relaxation.
This scheme was  recently proposed by  West {\it et al.} 
\cite{west09}.
Thus the road is paved for a much broader applicability of
optimized dynamic decoupling.

\begin{acknowledgments}
We are grateful for the financial support of the 
DFG in project UH 90/5-1.
\end{acknowledgments}

\appendix{}

\section{Effective Hamiltonian from the primary level of QDD}
Here we discuss in more detail how $\widetilde H^\mathrm{eff}(t)$
arises from the time evolution on the primary level in 
QDD. In particular, we discuss why this operator is
smooth or even analytic in $t$ although it arises from the integration
of $H(t)=D_0+D_1 F(t)$ where $F(t)$ is a switching function
which changes sign at the pulses of the primary level.

\begin{figure}[ht]
    \begin{center}
    \includegraphics[width=0.99\columnwidth,clip]{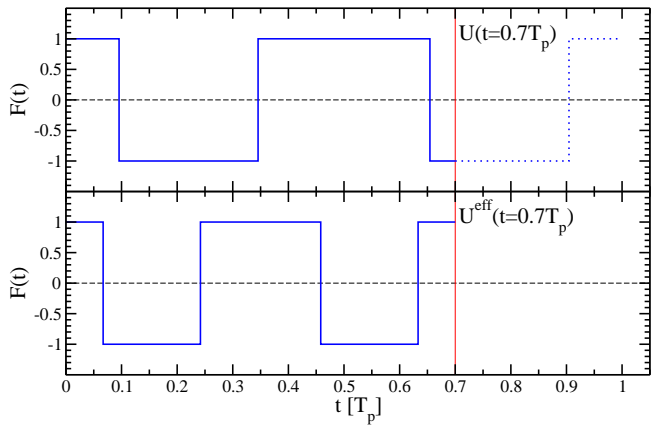}
    \end{center}
    \caption{(Color online) 
      Ranges of integration (range of solid curve) 
      and behavior of the switching
      function $F(t)$ for the calculation of $U(t)$ (upper panel) and of
      the effective $U^\mathrm{eff}(t)$ (lower panel).
      \label{fig:smooth}
}
\end{figure}

Let us denote by $U(t)$ the time evolution operator
on the primary level, i.e., within one of the intervals
of the secondary level, see Fig.\ \ref{fig:nesting}. It reads
\begin{equation}
\label{U-of-t1}
U(t)= {\cal T}\exp\big( -i\int_0^t H(t')dt' \big)
\end{equation}
and the domain of integration is illustrated
in the upper panel of Fig.\ \ref{fig:smooth}
by the range of the switching function $F(t)$ encountered.
If we define the dimensionless relative time $\tilde t:= t/T_\mathrm{p}$
and the corresponding Hamiltonian 
$H^\mathrm{rel}(\tilde t):= H(T_\mathrm{p} \tilde t)$ the
previous equation becomes
\begin{equation}
\label{U-of-t2}
U(t)=
{\cal \widetilde T}\exp\big( -iT_\mathrm{p}
\int_0^{t/T_\mathrm{p}} H^\mathrm{rel}(\tilde t)d\tilde t \big)
\end{equation}
where ${\cal \widetilde T}$ stands for the time ordering
according to the relative time $\tilde t$. Eq.\ \eqref{U-of-t2}
is given here for comparison with the subsequent effective
time evolution operator.

To consider what happens on the secondary level, we need
$U^\mathrm{eff}(t)$. Its defining property is
$U^\mathrm{eff}(T_\mathrm{p}) := U(T_\mathrm{p})$.
But if its argument is varied $T_\mathrm{p}\to t $ it is implied that the
primary sequence is scaled accordingly. That is its decisive
difference to $U(t)$ which is illustrated in
the lower panel of Fig.\ \ref{fig:smooth}. 
Note that this scaling is exactly what is done 
when the primary sequences are applied in each of
the time intervals of varying duration of the
secondary level, see Fig.\ \ref{fig:nesting}.

In a formula, $U^\mathrm{eff}(t)$  is given by
\begin{equation}
U^\mathrm{eff}(t)=
{\cal  \widetilde T}\exp\big( -i t
\int_0^{1} H^\mathrm{rel}(\tilde t)d\tilde t \big).
\end{equation}
From this equation it is obvious that $U^\mathrm{eff}(t)$
is generically smooth in the variable $t$. If the parts of the Hamiltonian are
bounded $U^\mathrm{eff}(t)$ is analytic because the
exponential is analytic.

The final step is to define the corresponding effective
Hamiltonian $\widetilde H^\mathrm{eff}(t)$.
As usual the Hamiltonian is retrieved from the time evolution
as its infinitesimal generator
\begin{equation}
\widetilde H^\mathrm{eff}(t) := i\left[\partial_t U^\mathrm{eff}(t)\right]
\left[ U^\mathrm{eff}(t)\right]^\dagger.
\end{equation}
If $U^\mathrm{eff}(t)$ is analytic, then $\widetilde H^\mathrm{eff}(t)$
is it as well. If $U^\mathrm{eff}(t)$ can be expanded
 up to and including the power $t^{N+1}$,  then $\widetilde H^\mathrm{eff}(t)$
can be expanded  up to and including the power $t^{N}$.


\end{document}